\documentclass{osa-article}

\journal{osac}

\articletype{Research Article}

\usepackage{lineno}
\usepackage{amsmath}
\usepackage{graphicx}
\usepackage{bm}
\usepackage{appendix}
\usepackage{caption}
\usepackage{multirow}
\usepackage{comment}

\begin{document}

\title{Focal-plane wavefront sensing with photonic lanterns II: numerical characterization and optimization}

\author{
Jonathan Lin,\authormark{1,*} 
Michael P. Fitzgerald,\authormark{1} 
Yinzi Xin,\authormark{2}
Yoo Jung Kim,\authormark{1}
Olivier Guyon,\authormark{3}
Sergio Leon-Saval,\authormark{4}
Barnaby Norris,\authormark{5}
Nemanja Jovanovic\authormark{2}
}

\address{\authormark{1} Physics \& Astronomy Department, University of California, Los Angeles (UCLA), 475 Portola Plaza, Los Angeles 90095, USA\\
\email{\authormark{*}jon880@astro.ucla.edu}}

\address{\authormark{2} Department of Astronomy, California Institute of Technology, Pasadena, CA, 91125, USA\\}
\address{\authormark{1} Physics \& Astronomy Department, University of California, Los Angeles (UCLA), 475 Portola Plaza, Los Angeles 90095, USA\\
\email{\authormark{*}jon880@astro.ucla.edu} 
\authormark{3} Department of Astronomy and Steward Observatory, The University of Arizona, 933 N. Cherry Ave., Tucson, AZ 85719, USA\\
\authormark{4} Sydney Astrophotonic Instrumentation Laboratory, School of Physics, The University of Sydney, Sydney, NSW 2006, Australia \\
\authormark{5} Sydney Institute for Astronomy, School of Physics, Physics Road, The University of Sydney, NSW 2006, Australia}


\begin{abstract}
We present numerical characterizations of the wavefront sensing performance for few-mode photonic lantern wavefront sensors (PLWFSs). These characterizations include calculations of throughput, control space, sensor linearity, and an estimate of maximum linear reconstruction range for standard and hybrid lanterns with 3 to 19 ports, at $\lambda = $ 1550 nm. We additionally consider the impact of beam-shaping optics and a charge-1 vortex mask, placed in the pupil plane. The former is motivated by the application of PLs to high-resolution spectroscopy, which could enable efficient injection into the spectrometer along with simultaneous focal-plane wavefront sensing; similarly, the latter is motivated by the application of PLs to vortex fiber nulling (VFN), which can simultaneously enable wavefront sensing and the nulling of on-axis starlight. Overall, we find that the PLWFS setups tested in this work exhibit good linearity out to $~\sim 0.25-0.5$ radians of RMS wavefront error (WFE). Meanwhile, we estimate the maximum amount of WFE that can be handled by these sensors, before the sensor response becomes degenerate, to be around $\sim 1-2$ radians RMS. In the future, we expect these limits can be pushed further by increasing the number of degrees of freedom, either by adopting higher-mode-count lanterns, dispersing lantern outputs, or separating polarizations. Lastly, we consider optimization strategies for the design of the PLWFS, which involve both modification of the lantern itself and the use of pre- and post-lantern optics like phase masks and interferometric beam recombiners. 
\end{abstract}

\section{Introduction}\label{sec:intro}

Photonic lanterns (PLs) are a type of optical and infrared waveguide which efficiently couple aberrated light into single-mode fibers (SMFs). In this work, we apply the PL as a focal-plane wavefront sensor, enabling the sensing and correction of non-common-path aberrations (NCPAs). While the photonic lantern wavefront sensor (PLWFS) is a relatively new concept, its capability has already been shown experimentally: for instance, \cite{Norris:20} demonstrated retrieval of the first 9 non-piston Zernike aberrations using a 19-port lantern and a neural net reconstruction scheme. We take a complementary approach, and provide an initial investigation into the PLWFS in the small aberration regime, using simpler reconstruction techniques motivated by analysis from first principles. In Paper I \cite{paper1}, we developed a mathematical framework for the PLWFS intensity response and derived linear and higher-order reconstruction models, as well as quantitative metrics for sensing performance. Our work focused on linear analysis of the PLWFS, more analogous to what is done in conventional adaptive optics (AO) control theory. The linear regime is also more relevant in scenarios where the PLWFS drives a dedicated sensing and correction stage, complementary to pupil-plane AO. This companion paper builds on Paper I by using numerical models to characterize the WFS properties of few-mode ``standard'' and ``hybrid'' PLs (see \S\ref{ap:a}) in the near infrared, the wavelength regime for upcoming instruments such as HISPEC and MODHIS \cite{Mawet:19:HISPEC,Mawet:22}. For simplicity, we neglect noise. While our models are undoubtedly idealizations, they will nonetheless help to determine manufacturing requirements for future lantern-based WFS systems, and will broaden our understanding on how the PLWFS architecture can be tuned in design and manufacture. 
\begin{figure}
    \centering
    \includegraphics[width=\textwidth]{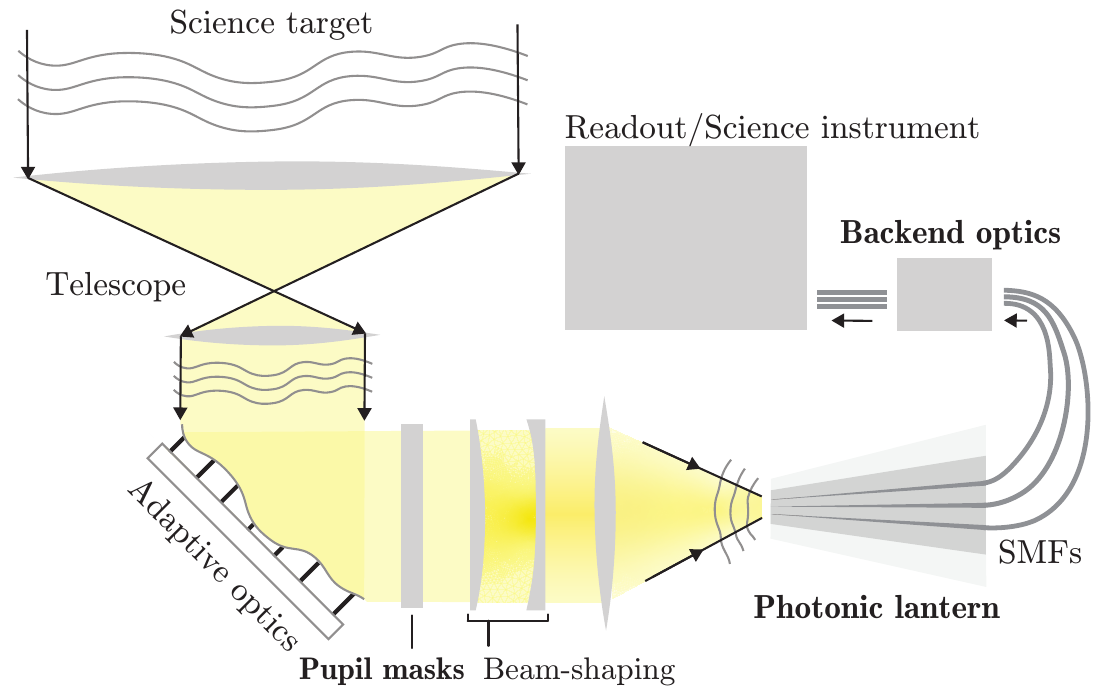}
    \caption{Diagram of a PLWFS, with the PL located in the focal plane of an AO-equipped telescope. The telescope's front-end AO may have an additional non-common-path WFS for first-stage correction (not shown). After first-stage correction, telescope light may pass through pupil-plane optics (phase masks and beam-shaping lenses) before being injected into the lantern; the coupled light can then be passed through ``backend optics'', our general term for any sort of device that operates on the single-mode PL outputs, before being routed to a detector or other science instrument. In this work, we also consider how the overall properties of the PLWFS may be optimized by modification of the three bolded components.}
    \label{fig:PL}
\end{figure}
\\\\
Figure \ref{fig:PL} presents an overview of the PLWFS setup assumed in this work, where a PL in the focal plane of an AO-equipped telescope couples light into an array of SMFs, which are eventually routed to a detector. For this basic configuration, where the lantern outputs are photometrically measured and polarization is neglected, an $N$-port PL has at most $N$ degrees of freedom. These $N$ degrees may be applied in different ways, for instance adding sensitivity to more modes in the linear regime, or increasing the reconstruction range of the sensor through non-linear techniques. In this work we restrict our analysis to few-moded lanterns with $N$ on the order of 10, which act as low-spatial order WFSs and are well-suited to sense NCPAs in a light-efficient manner. In principle, sensing is also possible for PLs with $N$ on the order of 100, though such lanterns have not yet been demonstrated in astronomy either numerically or experimentally; we leave this for future work. The number of degrees of freedom can also be increased for a few-port PL by adding optics which disperse its outputs spectrally and/or by polarization state, since PLs are both chromatic and polarization-dependent; these considerations are also left for future work.
\\\\
Beyond the benefits of focal-plane wavefront sensing, the PLWFS is additionally compelling because it can simultaneously serve applications including high-resolution spectroscopy and high-contrast imaging \cite{Trinh:13:GNOSIS,ellis:18:PRAXIS,Lin:21,Xin:22,kim:22}. To this end, beyond considering the PLWFS in isolation, we also consider performance in two promising non-WFS (and not necessarily exclusive) contexts: as a coupling device for SMF-fed, diffraction-limited spectrometers (e.g. \cite{Lin:21}), and as a platform for vortex fiber nulling (VFN; \cite{Ruane:19:VFN,Xin:22}). The former typically uses beam-shaping (or phase-induced amplitude apodization/PIAA) lenses to remap the focal-plane amplitude distribution of light in order to maximize coupling into an SMF \cite{Jovanovic:17}. Beam-shaping optics combined with a hybrid PL work in a similar way, routing the bulk of the light into a single-mode ``science" fiber, while coupling extra light into non-science ports which can then be used for wavefront sensing \cite{hybrid}. The second application we consider is VFN, a high-contrast imaging technique where telescope light is passed through a helical phase mask and then injected into an SMF. Since the resulting PSF for an on-axis source is orthogonal to the SMF's fundamental mode, at least in the absence of aberrations, on-axis starlight is nulled while any off-axis planet light is partially coupled \cite{Ruane:19:VFN}. In place of an SMF, VFN can also use a PL, given that at least one of the the PL's ``principal modes'' (see \S\ref{ap:a}) is invariant under a partial rotation. This criterion is clearly satisfied by hybrid lanterns, which have the rotationally invariant fundamental fiber mode as a principal mode, but can also be satisfied by standard lanterns with certain symmetries. Extending VFN to use a PL instead of an SMF can enable nulling in conjunction with simultaneous wavefront control; alternatively, combining a VFN with a fully mode-selective lantern (see \ref{ap:a}) may also boost achievable sensitivity and tighten inner working angle \cite{Xin:22}.
\\\\
Our second focus, after characterizations, is to find ways to optimize the PLWFS. Strategies are divided into three categories: through physical modification of the lantern itself, via geometric properties such as taper length or core arrangement; through ``frontend'' modification of light upstream of the lantern, via pupil-plane masks and DM control; and through ``backend'' modification of the light downstream of the lantern, via coherent beam recombination. Optics corresponding to each strategy are bolded in Figure \ref{fig:PL}. Frontend and backend optimizations are particularly important because they are independent of the PL, giving pathways through which the PLWFS can be tuned even after the PL has been fabricated. In this work, we present initial investigations into all three strategies. Overall, the questions we address in this work can be divided into two areas --- characterization and optimization  --- and are organized in Table \ref{tab:contents}.
In \S\ref{sec:method}, we outline our numerical model. In \S\ref{sec:results}, we show the results of this model, applied to an ensemble of PLWFS configurations, while in \S\ref{sec:opt} we discuss further pathways through which the sensing properties of the PLWFS can be optimized.

\begin{center}
\begin{table}
\centering
\caption{PLWFS questions} \label{tab:contents}
\begin{tabular}{|| p{2.4cm} | p{8cm} | p{1.4cm}||} 
\hline 
Area & Question & Section(s) \\
\hline\hline
\multirow{4}{*}{Characterization} & What are the useful metrics for benchmarking PLWFSs? & \S\ref{ssec:metric}\\
\cline{2-3}
& What are the impacts of beam-shaping (PIAA) optics? & \S\ref{ssec:beamshaping} \\
\cline{2-3} 
& What are the impacts of vortex masks? & \S\ref{ssec:vortex}\\
\cline{2-3}
& What approximate wavefront sensing performance might we expect from 3-, 6-, 10-, 12-, and 19-port PLs? & \S\ref{ssec:allresults}\\
\hline
\multirow{4}{2.5cm}{Design optimization} & Can we control sensing properties by tuning PL design parameters such as taper length and core spacing? & \S\ref{ssec:length}-\S\ref{ssec:coregeom}\\
\cline{2-3}
& What {\it pre-lantern} optics can we add to improve the PLWFS’s sensing properties? & \S\ref{ssec:masks}\\
\cline{2-3}
& What {\it post-lantern} optics can we add to improve the PLWFS’s sensing properties? & \S\ref{ssec:beamcombining}\\
\hline
\end{tabular}
\end{table}
\end{center}

\section{Simulations}\label{sec:method}
We numerically model the PLWFS and characterize its ability to sense low-order phase-only aberrations, in monochromatic light ($\lambda = 1.55 \,\upmu$m). While pure monochromatic light in unrealistic, we anecdotally expect PLs to perform similarly out to bandwidths of perhaps 100 nm: this is roughly the scale of variation we observe in the spectral traces of the dispersed 3- and 19-port PLs installed in the SCExAO/Subaru testbed \cite{Lozi:20}, when injecting a broadband source. When simulating the PLWFS, we consider two separate phase aberration bases: a fixed basis consisting of the first five non-piston Zernike modes; and the ``control modes'', the right-singular vectors of the linear response matrix measured about zero WFE, which vary from lantern to lantern. While sensing is often done with the latter type of basis in conventional AO, the former is useful because it provides a common reference against which different PLWFS configurations can be compared. 
\\\\
The PLWFS model has three components: a telescope model, which takes in an incident wavefront and returns a complex-valued focal plane electric field; a numerical beam propagator, which takes both an incident focal-plane electric field and a waveguide geometry, and returns the electric field at the waveguide output (neglecting polarization); and a wavefront reconstructor, derived from our analysis in \cite{paper1}. Section \ref{ssec:tele} gives an overview of our telescope model. Lantern propagation is handled using the adaptive mesh code Lightbeam \cite{lightbeam}, using a coarse resolution of $0.5$ $\upmu$m in $(x,y)$, and a $z$ stepsize of 1-5 $\upmu$m (depending on lantern length); for more details on the lantern propagation and wavefront reconstruction components of our numerical model, refer to \cite{paper1}. In \S\ref{ssec:param}, we cover the different PLWFS configurations considered in this work, and in \S\ref{ssec:metric}, we define the metrics used to benchmark the sensing performance of those configurations.

\subsection{Telescope simulation} \label{ssec:tele}
Propagation through telescope optics is handled using HCIPy, a Python-based high-contrast imaging package \cite{hcipy}. Simulations are monochromatic at a wavelength of 1.55 $\upmu$m. We additionally assume a 10 m circular, unobstructed aperture; the focal ratio of the system is optimized to maximize total throughput of an unaberrated wavefront. Before wavefronts are propagated to the focal plane, we can simulate an arbitrary phase mask $\bm{\phi}$ by multiplying the pupil-plane electric field by $\exp(i\bm{\phi})$. These phase masks are relevant for certain high-contrast use cases (e.g. vortex masks for VFN), but can also be designed to optimize WFS performance, as discussed later in \S\ref{sec:opt}. We additionally implement beam-shaping lenses, using HCIPy's Fresnel propagator; the lens profiles themselves are numerically calculated following \cite{Guyon:03:PIAA}. Our particular implementation of beam-shaping promotes a Gaussian focal-plane amplitude distribution in the absence of wavefront error, similar to what is used in SMF injection. Lastly, pupil-to-focal plane propagation is handled via HCIPy's Fraunhofer propagator. 

\subsection{PL parameter space}\label{ssec:param}
All PLs simulated in this work obey the following assumptions. First, we assume that every PL tapers uniformly and linearly, and that cross-sections of the cores and overall cladding remain perfectly circular through the taper. Additionally, all PLs taper by a factor of 8 from entrance to exit (i.e. the geometry of the small end is the same as the large end, but scaled down by 8$\times$), with cores spaced in the cladding to reflect the geometries produced when constructing lanterns from bundles of uniformly-sized SMFs; for an initial investigation into different core spacings, see \S\ref{ssec:coregeom}. The cladding index is set to 1.444, corresponding to fused silica at $\lambda = $ 1.55 $\upmu$m, while the jacket-cladding contrast is set to $5.5\times10^{-3}$, corresponding to a fluorine-based glass. The core index is set so that the mode field diameter is $\sim$7.5 $\upmu$m, matching OFS ClearLite 980 16 fiber.
\begin{figure}
    \centering
    \includegraphics[width=\textwidth]{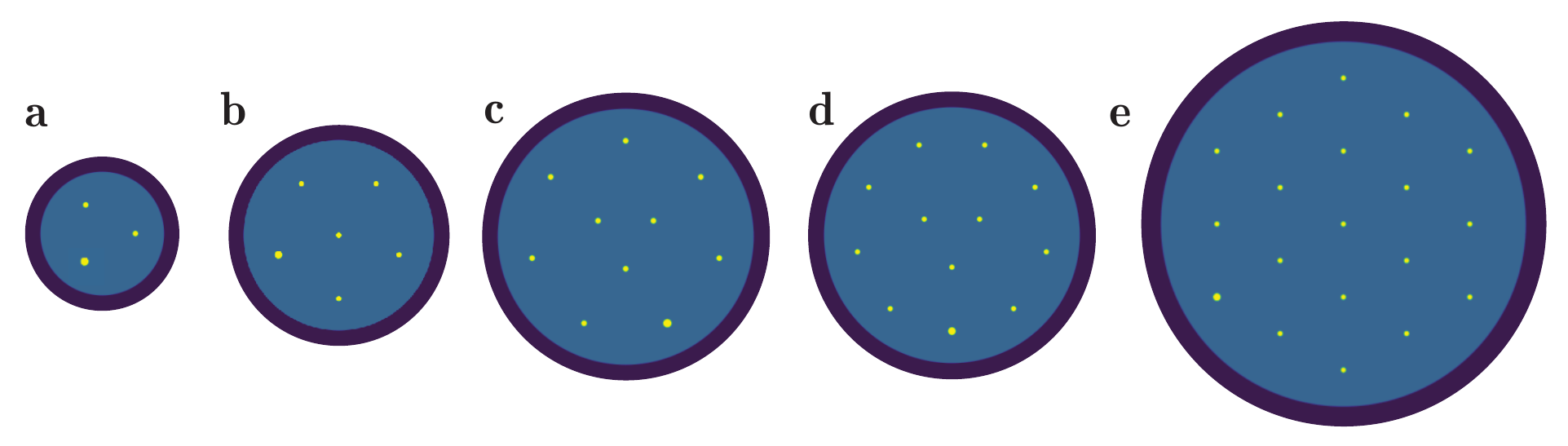}
    \caption{Refractive index cross-sections for tested PLs. Panels a, b, c, d, and e show geometries for hybrid 3-, 6-, 10-, 12- and 19-port hybrid lanterns, respectively. Non-selective variants are not shown for brevity; they are similar in structure to the hybrid lanterns but with uniform core size. }
    \label{fig:geoms}
\end{figure}
\\\\
Under the above assumptions, we analyze five sizes of PL: 3-, 6-, 10-, 12- and 19-port, whose output-end geometries are shown in Figure \ref{fig:geoms}. Each lantern size has two variants: a ``standard,'' non-selective variant where all SMF cores have the same diameter ($4.4$ $\upmu$m), and a hybrid variant where one core is made 2 $\upmu$m micron larger in diameter to select out the LP$_{01}$ mode. For all hybrid PLs, the selective core is placed as far off-center as possible, in order to break mode structure symmetries and minimize cross-talk. This choice is especially important for the 6-port hybrid lantern, which would be insensitive to defocus if the central core was made selective. Notably, the hybrid 3-port lantern is technically ``mode-group-selective'' (see \ref{ap:a}). By a similar argument to what was laid out for fully mode-selective lanterns in \cite{paper1}, we also expect mode-group-selective lanterns to have poor WFS performance without help from devices like backend photonic integrated circuits (PICs). Regardless, we include this lantern in our simulations for completeness. 
\\\\
We additionally optimize the length of the transition region for standard 3- and 6-port lanterns, and the hybrid 6-port lantern, to improve the linearity of their intensity response to phase aberrations. This optimization is done by setting the PL to an initial length (40 mm for the 6-port hybrid lantern and 20 mm for the rest) and entrance diameter (12 $\upmu$m for the 3-port, 20 $\upmu$m for the 6-port), and then gradually shortening the lantern, as if material is being cleaved from the lantern entrance, while computing the range in aberration amplitude for which linear reconstruction is accurate to within 0.1 radians RMS, for each aberration mode. The product of these ranges gives a rough metric for linearity (for a more precise but computationally expensive metric, refer to ``linear radius'', \S\ref{ssec:metric}). As the lantern is shortened, and the entrance diameter increases, we also repeatedly apply focal ratio optimizations to keep the coupling of the PSF into the PL maximal. For simplicity, these optimizations were done in the absence of beam-shaping optics and vortex masks; a more in-depth treatment on the interaction between length optimization and different lantern optics is outside the scope of this work. Similarly, we opted not to apply this optimization to the larger 10-, 12-, and 19-port lanterns since the added complexity of these lanterns makes optimization less straightforward: at least anecdotally, gains in reconstruction range in one phase aberration mode typically come at the cost of another, and overall the effect of length optimization on WFS properties seems to be lesser for larger lanterns with more outputs. The 10- and 12-port standard lanterns have fixed lengths of 20 mm, while the hybrid variants have lengths of 60 mm. The standard 19-port lantern has a fixed length of 60 mm while the hybrid has a length of 100 mm. The hybrid lanterns are made longer to reduce cross-talk and improve their mode-selective properties. Entrance diameters are 24 $\upmu$m, 26 $\upmu$m, and 37.6 $\upmu$m  for the 10-, 12-, and 19-port lanterns, respectively. 
\\\\
Finally, for each lantern geometry mentioned above, we additionally test the impact of a charge-1 vortex mask of the system, implemented as a helical phase ramp in the pupil plane. As a caveat, the 3-, 10-, and 12-port lanterns with the vortex mask do not function as nullers, but we include their results anyways to assess the impact the vortex mask has on wavefront sensing. We also consider the impact of beam-shaping, but only for the 6-port hybrid PL: we expect beam-shaping optics (which apply an amplitude apodization) to impact WFS properties in a less drastic and more consistent manner than vortex masks, so that any specific results will be qualitatively generalizable to other lanterns. 

\subsection{Performance metrics}\label{ssec:metric}
We consider the following five metrics:
\paragraph{Throughput} We define throughput as the fraction of power of the incident electric field that makes it through the entire telescope + PLWFS system. All optical losses are assumed to occur during coupling into or propagation through the PL; other losses (e.g. from imperfect telescope optics) are neglected. 
\paragraph{Sensed low-order Zernikes} We consider, for each PLWFS configuration, which subset of the first five non-piston Zernikes (tip/tilt, defocus, and astigmatisms) can be sensed simultaneously - these low-spatial-frequency modes will likely contain the bulk of the power in both atmospheric and instrumental WFE, and form a common basis for later comparisons.
\paragraph{Rank}
The maximum number of modes that can be sensed using a linear reconstructor is equal to the rank of the interaction matrix (e.g. $B$ in equation 5 of \cite{paper1}). To measure rank numerically for each PLWFS configuration, we form the interaction matrix for the first 30 non-piston Zernikes, and compute the singular value decomposition, which produces the control modes of the sensor. We then count the number of modes with singular value larger than 0.05 simulated flux unit/radian (flux units are normalized to set the power of incoming wavefronts to one). 
\paragraph{Linear radius $r_L$} When considering only a single aberration mode, we define the linear range as the interval in mode amplitude within which linear reconstruction is accurate to 0.1 radians RMS. When sensing $N$ modes simultaneously, phase aberrations become an $N$-length vector of mode amplitudes, and the region where linear reconstruction is accurate becomes $N$-dimensional. We estimate the volume of this region using Monte Carlo integration. Approximating this $N$D region as a ball then gives the ``linear radius'' $r_L$ in units of radians RMS. Aberrations with total RMS WFE less than this radius can generally be reconstructed with good accuracy using the linear model. In principle, non-linear models and closed-loop control should be able to perform wavefront reconstruction beyond this radius. 
\\\\
Because $r_L$ depends on the phase aberration modes being sensed, we compute it for both the low-order Zernike basis and control mode basis.

\paragraph{Degenerate radius $r_D$} Following \S3.4 in \cite{paper1}, the degenerate radius $r_D$ estimates the maximum amount of RMS WFE an aberration can have before the process of phase retrieval becomes ill-posed; see also \S\ref{ap:deg} for a minor clarification. Beyond this radius, a WFS may map distinct mode amplitude vectors (each corresponding to a different phase aberration) to the same intensity response, making the propagation process non-invertible without additional regularization. This metric is an approximation, designed to be easily calculable numerically, and the presence of degeneracies do not necessarily impose a fundamental barrier to WFS if the volume of phase space within which degeneracies occur is small. Degeneracy might also be lifted by spectral dispersion of the lantern outputs.
\\\\
Like for $r_L$, we compute $r_D$ for both the low-order Zernike basis and for control mode basis.

\section{Results}\label{sec:results}
In this section, we use our numerical model to compute the above metrics for various PLWFS configurations, additionally considering performance in two contexts: injection for diffraction-limited spectrometers (\S\ref{ssec:beamshaping}), and VFN (\S\ref{ssec:vortex}). In \S\ref{ssec:allresults}, we review the performance of all tested configurations.

\subsection{Beam-shaping}\label{ssec:beamshaping}
\begin{figure}
    \centering
    \includegraphics[width=\textwidth]{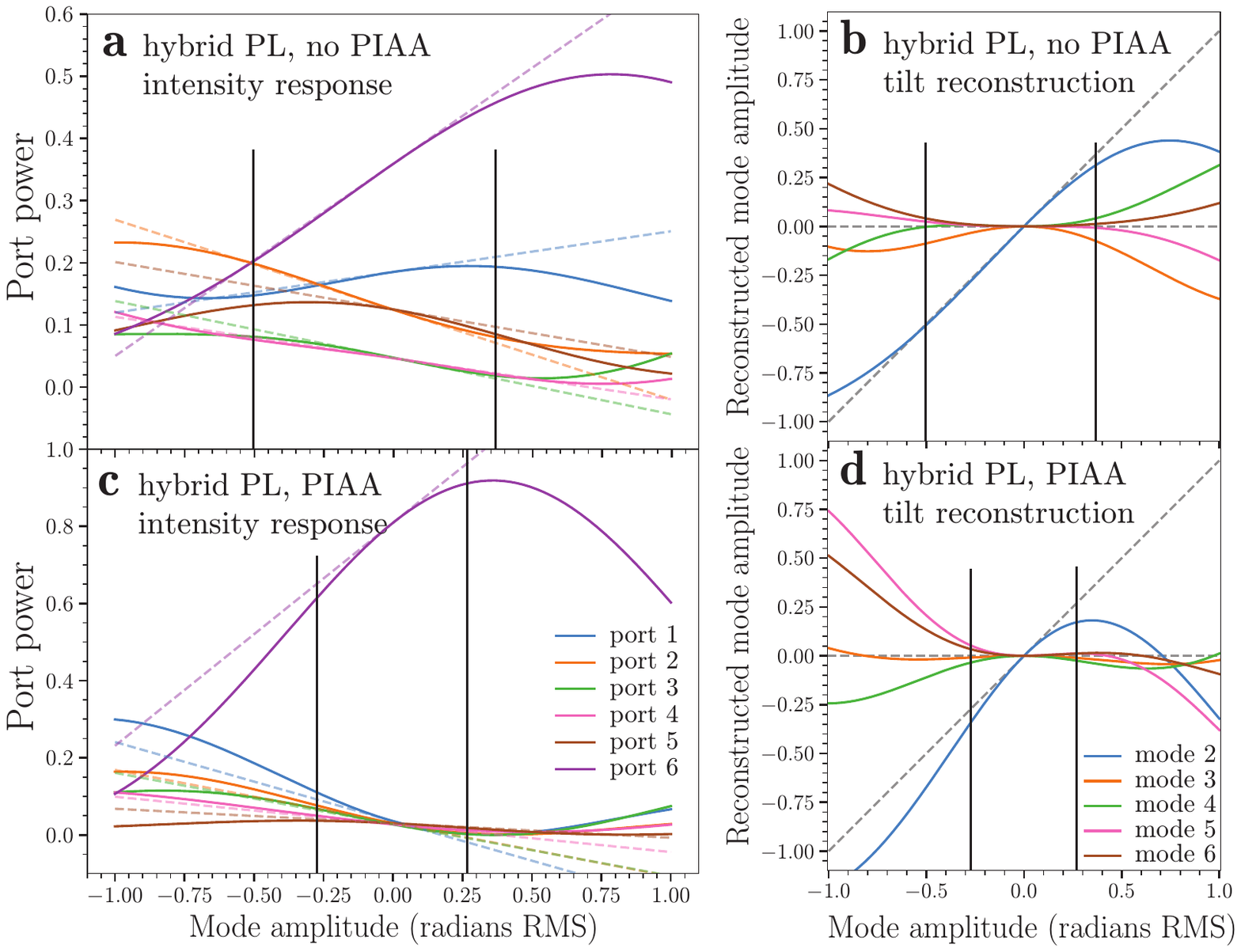}
    \caption{Panel a: intensity response of the hybrid 6-port lantern in the presence of tilt, Zernike mode 2, without beam-shaping optics. Dashed, colored lines show the linear approximation to the WFS response, while vertical black bars denote the linear range. Panel b: linear reconstruction of the tilt mode for the same lantern. Under perfect reconstruction, the trace corresponding to the scanned mode should follow the line $y=x$ (diagonal dashed line) while all other traces should follow $y=0$ (horizontal dashed line). Panels a and b are adapted from \cite{paper1}. Panels c, d: same as Panels a and b but for a hybrid 6-port lantern with beam-shaping (PIAA optics). The presence of said optics boost coupling into port 6, the science port. As a tradeoff, we see that beam-shaping reduces the linear range of the sensor by around 30--40\%. }
    \label{fig:piaa}
\end{figure}

Here, we simulate the interaction between beam-shaping optics and PLs. These tests are motivated by SMF-fed, high-resolution spectrometry, where it is desirable from a detector-usage perspective to couple as much light as possible into a single fiber. Accordingly, this section only considers hybrid lanterns, which can couple the bulk of light into their selective port, at least in the absence of wavefront aberrations. In contrast, standard lanterns have a tendency to distribute light more evenly among their output ports, all of which would need to be routed to the spectrometer. Mode-selective lanterns are also not well-suited for this application: though they can couple the bulk of light into a single port like hybrid PLs, such lanterns by themselves lack WFSing capability \cite{paper1}. The combination of Gaussian beam-shaping optics and hybrid PLs should boost single-port coupling even further, since such optics promote a Gaussian amplitude distribution for incoming light, which matches the approximately Gaussian-like principal mode of hybrid lanterns. 
\\\\
Simulations recover these expectations. Panel a of Figure \ref{fig:piaa} shows the intensity response of a 6-port hybrid lantern without beam-shaping optics, in the presence of tilt (Zernike mode 2), while Panel c plots the same for a hybrid lantern with beam-shaping optics. Comparing the panels, we see that beam-shaping more than doubles coupling into port 6 (the selective port) from 35\% to 80\%. This increase in coupling to the selective port comes at the cost of sensing range: with less light overall in the non-selective lantern ports, the PLWFS behaves more non-linearly.  
Our results are reiterated in Figures \ref{fig:piaa}b and d, which show linearly-reconstructed tilt amplitude for the sensing configurations without and with beam-shaping, respectively. Beam-shaping reduces the linear reconstruction range by around 40\%. We obtain similar results when scanning over defocus and astigmatism. 

\subsection{Vortex masks}\label{ssec:vortex}
\begin{figure}
    \centering
    \includegraphics[width=\textwidth]{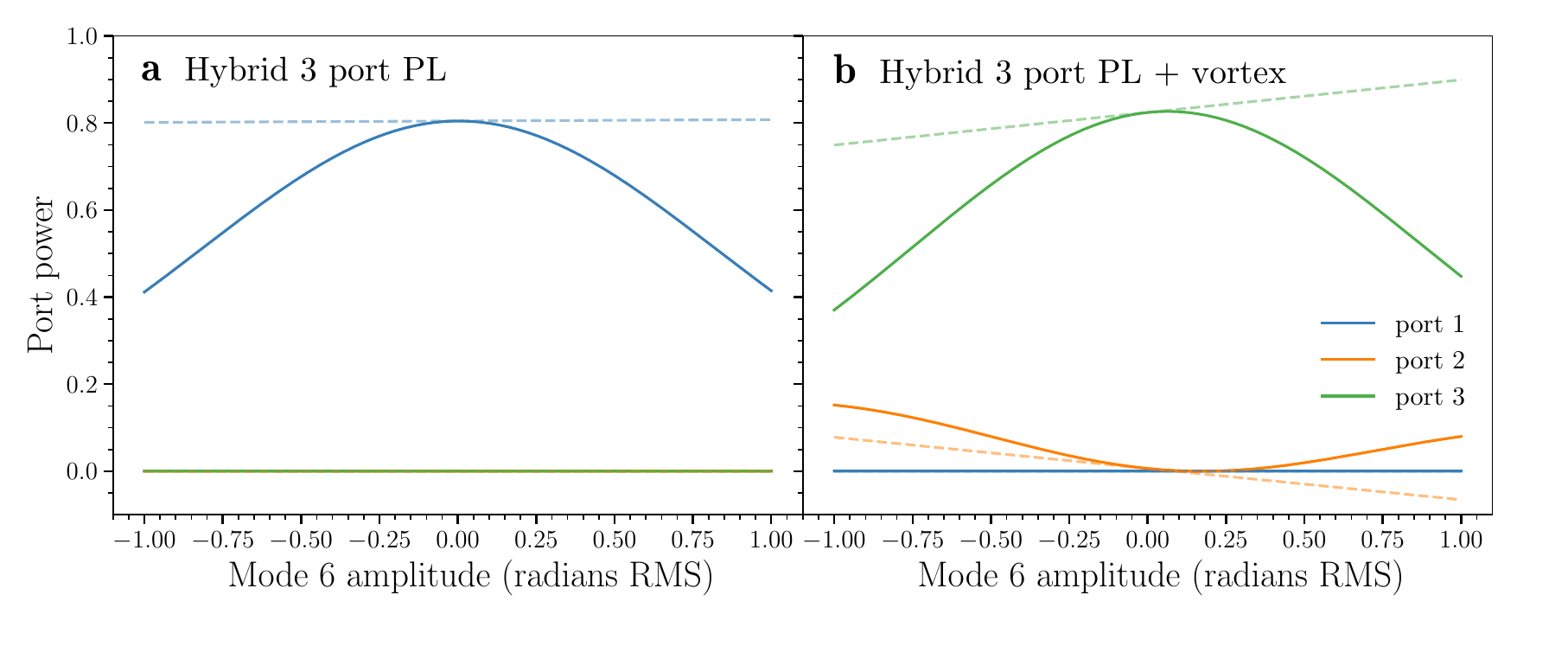}
    \caption{Intensity response of a hybrid 3-port PL (a) without a pupil-plane vortex mask and (b) with a pupil-plane vortex mask, in the presence of Zernike mode 6. Dashed lines show the slopes of the intensity response in each port, about the origin. Without the vortex mask, only port 1 (the selective port) has non-zero flux, and the response in this port is completely symmetric, so sensing cannot be done. With the vortex, the selective port is nulled, and the symmetry in the intensity response is ultimately broken (albeit only slightly).}
    \label{fig:vortex}
\end{figure}

The impact of a charge-1 vortex mask on sensing performance depends strongly on the azimuthal symmetry of the principal lantern modes (see \cite{paper1}, Section \S2.1), which in turn depends on factors like core arrangement and lantern length. One of the main impacts of a vortex mask is to alter the sensed set of aberration modes. The clearest example is the hybrid 3-port PL, which cannot sense any modes without the vortex mask, but can sense Zernike mode 6 with the mask in. This effect is shown in Figure \ref{fig:vortex}, which plots intensity responses in the presence of Zernike mode 6 for the 3-port hybrid lantern, in the absence and presence of a pupil-plane vortex mask: the vortex mask makes the intensity response of the lantern asymmetric, so that the response of the system is no longer degenerate in the linear regime.
\\\\
While the vortex mask is beneficial in this particular scenario, the effect is not consistent. In fact, the vortex mask can occasionally remove sensitivity to a particular mode (for instance, defocus with the 6-port PL); vortex masks can also impact linear and degenerate radii both positively and negatively. In the next section we present tabulated results that show the effect of the vortex mask on case-by-case basis.

\subsection{Characterization of PLWFS configurations}\label{ssec:allresults}
Here, we evaluate the five metrics detailed in Section \S\ref{ssec:metric} for all tested PLWFS configurations. Results are summarized in Table \ref{tab:standard} for standard PLs, and Table \ref{tab:hybrid} for hybrid PLs. We also outline some qualitative trends that can be seen in the results, below. 
\begin{itemize}
    \item Consistent with \cite{Lin:21}, we find that throughput increases with the number of ports.
    \item  Rank also increases with number of ports, although no configuration is able to sense as many modes as it has ports. We believe this is because one of the PLWFS's degrees of freedom is taken by a piston-like mode, corresponding to the LP01 mode backpropagated to the pupil plane. See further discussion in \ref{ssec:PLWFSchar}.
    \item $r_D$ drops significantly when switching from the low-order Zernike basis to the control mode basis, when the number of control modes exceed the number of sensed Zernikes. In other words, the ability to sense more modes comes at the cost of sensor range.
    \item For standard lanterns, the degenerate radius in control mode basis is typically improved by the vortex mask; there is no such trend for hybrid lanterns. One explanation is that the vortex mask introduces an asymmetry, where the intensity response of a standard lantern was before symmetric.
    \item Considering the 10-, 12-, and 19-port PLs, the linear radius in the low-order Zernike basis increases as the number of ports increases. This is expected due to the increased degrees of freedom. The 3- and 6-port lanterns break this trend because they were already optimized for linearity. 
    \item The vortex mask often can add sensitivity to a certain Zernike mode (see the hybrid 3-port PL, and both 10-port PLs), but occasionally removes sensitivity (see both 6-port PLs). 
    \item Configurations using hybrid PLs typically have a smaller linear radius compared to their standard PL counterparts; this is consistent with our result from \S5.1 in \cite{paper1} that hybrid lanterns become non-linear more quickly than their standard counterparts. Notable exceptions are the the 10- and 12-port lanterns, without the vortex mask.
    \item Hybrid lanterns also typically lose rank compared to their standard counterparts. This is unsurprising since the selective port is not useful for wavefront sensing. As an extreme case, the hybrid 3-port lantern has a completely symmetric intensity response for all aberration modes, and thus has both 0 rank and 0 degenerate radius.
    \item Degenerate radius is typically large by a factor of $\gtrsim 2$, motivating non-linear reconstruction techniques.
\end{itemize}
Finally, we note that in a few cases, the degenerate radius is actually less than the linear radius, when intuitively we might expect the opposite to always hold. This unintuitive behavior is clearest with the standard 19-port lantern, which in control mode basis has $r_L=0.30$ and $r_D=0.21$. The discrepancy is an artifact of how $r_L$ and $r_D$ are defined: recall that $r_L$ is an averaged quantity, but $r_D$ is not. Scenarios where $r_D<r_L$ occur when linearity is poor in one mode but not the rest: then that mode's nonlinearity will negatively impact $r_D$ more than $r_L$.
\\\\
Our results establish a rough estimate of PLWFS capability, with an upper limit of around 1--2 radians in RMS WFE before the onset of degeneracies, and an upper limit for linear reconstruction of up to 0.25-0.5 radians, depending on the choice of mode basis. Therefore, in cases where input aberrations are larger, the PLWFS may better suited in a secondary control loop behind a conventional pupil-plane wavefront control system, which can have a larger reconstruction range at the expense of non-common path. However, it remains to be seen if real photonic lanterns, which deviate from the ideal lanterns simulated in this work, will perform better or worse, though the qualitative trends observed in our modelling should be relatively robust. We also emphasize that our numbers are not yet fundamental: increasing the number of ports, exploiting polarization, and dispersing the lantern outputs are all future pathways to improving wavefront sensing capability. Beyond increasing the number of degrees of freedom, we can also make progress by altering the lantern structure and using additional optics. We discuss such prospects in the next section.

\begin{center}
\begin{table}
\centering
\caption{Performance of standard PLs}
\begin{tabular}{||c c c c c c c c||} 
 \hline
 Config. & Throughput   & Sensed ZMs & ZM $r_L$ & ZM $r_D$ & CM Rank & CM $r_L$ & CM $r_D$ \\ [0.5ex] 
 \hline\hline
 3 & 0.815 & 2-3 & 0.52 & 1.0 &2 & 0.54 & 2.8 \\ 
 \hline
 6 & 0.879  & 2-6 & 0.46 & 0.86 & 5& 0.47 & 0.92\\
 \hline
 10 & 0.890& 2-5 & 0.19 & 2.9  & 8 & 0.18 & 0.41 \\
 \hline
 12 & 0.895  & 2-6 & 0.27 & 2.0 & 11& 0.28 & 0.25 \\
 \hline
 19 & 0.921 & 2-6 & 0.49 & 2.3 & 15 & 0.30 & 0.21 \\
 \hline
 3V & 0.843  & 2-3 & 0.46 & 1.9& 2 & 0.51 & 0.88 \\
 \hline
 6V & 0.877 & 2-3,5-6 & 0.44 & 1.0 & 4 & 0.45 & 1.5\\
 \hline
 10V & 0.883 & 2-6 & 0.40 & 2.1 & 9 & 0.39 & 0.90 \\
 \hline
 12V & 0.883 & 2-6 & 0.47 & 2.3 & 11 & 0.43 & 0.97 \\
 \hline
 19V & 0.894 & 2-6 & 0.52 & 2.0 & 15 & 0.30 & 0.43\\ 
 [1ex] 
 \hline
\end{tabular}
\\[10pt]
\caption*{Throughput, rank, sensed Zernikes (out of modes 2-6), linear radius $r_L$, and degenerate radius $r_D$ for standard 3-, 6-, 10-, 12-, and 19- port lanterns, with and without a vortex mask in the pupil plane. Lantern configurations are identified in the first column by their number of ports; configurations with vortex masks are additionally marked with a capital ``V''. Linear and degenerate radius are presented for both a low-order Zernike mode (``ZM'') basis, formed from whichever of the first five non-piston Zernikes the lantern configuration can sense, and control mode (``CM'') basis.}
 \label{tab:standard}
\end{table}
\end{center}

\begin{center}
\begin{table}
\centering
\caption{Performance of hybrid PLs}
\begin{tabular}{||c c c c c c c c||} 
 \hline
 Config. & Throughput  & Sensed ZMs & ZM $r_L$& ZM $r_D$ & CM Rank & CM $r_L$& CM $r_D$ \\[0.5ex] 
 \hline\hline
 3 & 0.805 & None & 0 & 0 &0 & 0 & 0 \\
 \hline
 6 & 0.888 & 2-6 & 0.43 & 0.84 & 5 & 0.41 & 0.44\\
 \hline
 10 & 0.893 & 2-3,5-6 & 0.29 & 2.7 & 7 & 0.25 & 0.56 \\
 \hline
 12 & 0.898 & 2-6 & 0.39 & 2.3 & 10& 0.24 & 0.45 \\
 \hline
 19 & 0.921 & 2-6 & 0.39 & 3.1 & 13 & 0.30 & 0.87 \\
 \hline
 3V & 0.831 & 6 & 0.13 & - & 1& 0.13 & - \\
 \hline
 6V & 0.884 & 2-3,5-6 & 0.36 & 1.3 & 4& 0.37 & 1.5 \\
 \hline
 10V & 0.887 & 2-6 & 0.33 & 1.6 & 9& 0.26 & 0.71 \\
 \hline
 12V & 0.888  & 2-6 & 0.37 & 2.1 & 10& 0.29 & 0.54 \\
 \hline
 19V & 0.889  & 2-6 & 0.50 & 2.9 & 15& 0.27 & 0.77 \\ [1ex] 
 \hline
\end{tabular}
\\[10pt]
\caption*{Throughput, rank, sensed Zernikes (out of modes 2-6), linear radius $r_L$, and degenerate radius $r_D$ for hybrid 3-, 6-, 10-, 12-, and 19- port lanterns, with and without a vortex mask in the pupil plane, analogous to Table \ref{tab:standard}. The empty cells for the 3V configuration indicate that no degeneracies could be found, and that the system only becomes degenerate as aberration amplitude goes to infinity and flux in all outputs goes to 0.  }
 \label{tab:hybrid}
\end{table}
\end{center}

\section{Design optimization}\label{sec:opt}
In this section we consider how the wavefront sensing properties of the PLWFS can be improved. One strategy is to modify the structure of the lantern itself; however, optimizations of the optics before and after the PL are also important. Understanding of latter methods will enable more ways to fine-tune the properties of the PLWFS, some of which might be impossible to do through modification of the lantern alone. As a secondary motivation, ``optimal'' PL designs may not always be feasible to manufacture, but deviations from design specifications may be corrected by pre- and post-lantern signal processing.
\\\\
We first provide an initial investigation in optimization of the lantern itself, by changing a lantern's taper length and the radial spacing of its cores (\S\ref{ssec:length}--S\ref{ssec:coregeom}). We then consider how PLWFS performance may be further improved through pre-lantern phase masks (\S\ref{ssec:masks}), and in post-lantern beam recombination (\S\ref{ssec:beamcombining}).

\subsection{Lantern length}\label{ssec:length}
\begin{figure}
    \centering
    \includegraphics[width=\textwidth]{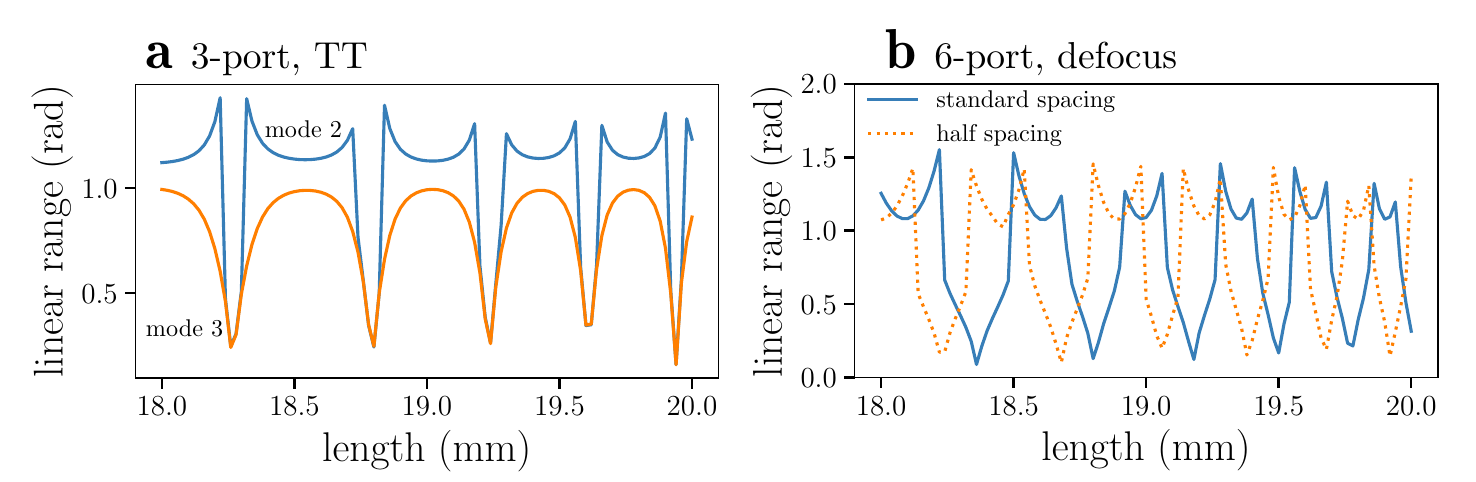}
    \caption{Panel a: Oscillatory behavior of tip (Zernike mode 2) and tilt (Zernike mode 3) reconstruction range for a standard 3-port lantern, as a function of lantern length. These oscillations occur over a period of around 300 $\upmu$m. Panel b: Oscillatory behavior of defocus (Zernike mode 4) reconstruction range for a non-mode-selective 6-port lantern, as a function of lantern length. Curves for two lanterns are shown: the first has ``standard'' radial spacing, where the outer single-mode cores are placed two-thirds along the way from the lantern's center to the cladding-jacket interface.  The second lantern places the outer lantern cores halfway between the lantern's center and the cladding-jacket interface. This change in radial spacing displaces the defocus reconstruction oscillations.  }
    \label{fig:osc}
\end{figure}

One way to optimize a PL for wavefront sensing is to tune the lantern's length (``taper length'') to maximize a sensing metric such as linear range. As an example, we simulate the effect lantern length has on linear reconstruction range for the standard 3-port PL specified in \S\ref{ssec:param}. Figure \ref{fig:osc}a tracks the linear reconstruction range for the tip and tilt modes of the PL as a function of length. Note that we alter length while holding the tapering angle constant, as if cleaving material from the lantern entrance. While the range of lengths that we tested is narrow enough to not alter the number of modes supported at the PL entrance, length also impacts the size of the entrance, requiring re-optimization of our telescope model's F\#. For our chosen PL, we find that the linear reconstruction ranges for the tip and tilt modes oscillate with length, at a period of around 300 $\upmu$m, likely due to beating between modes with different propagation constants. This oscillation period has real-word ramifications: in particular, it sets restrictions in the manufacturing tolerances for taper length, which must be comparatively small in order to enable control over the lantern's sensing properties.
\\\\
Figure \ref{fig:osc}b (blue curve) shows similar oscillations in linear reconstruction range for the standard 6-port lantern, specifically for the defocus mode. These oscillations are reminiscent of a phenomenon discussed in \cite{paper1}, which showed that defocus {\it sensitivity} (not reconstruction range, though the two metrics are related) peaks when the LP$_{01}$ and LP$_{02}$ components of the central lantern mode (i.e. the mode corresponding to the central port of the lantern) are 90$^\circ$ out of phase. In turn, oscillations in relative phase (which would lead to alternation in defocus reconstruction range, as seen in Figure \ref{fig:osc}b) might occur as a function of lantern length, since different modes travel with a different effective refractive index through the lantern.
\\\\
While oscillations in reconstruction range also occur for higher mode-count lanterns, the addition of more lantern modes complicates the optimization process. Different sets of modes within may oscillate at different periods, so that gains in for one mode are more likely to come at the cost of another. We leave this for future work.

\subsection{Core arrangement}\label{ssec:coregeom}
The geometric arrangement of single-mode cores within a standard PL can also strongly impact wavefront sensing ability \cite{Fontaine:12}. For example, prior work in the telecommunications field has shown that a PL's ``rings'' (sets of cores that are the same radial distance from the PL's center) should contain odd numbers of cores in order to break symmetries in the lantern mode structure \cite{Davenport:21}. In this section, we show that the radial spacing between rings also impacts wavefront sensing, using the standard 6-port lantern (assumed to be fabricated by tapering a bundle of SMFs) as an example. For a PL constructed from uniformly-sized SMFs, the outer ring of cores will be located at a distance of $2/3 \, r_{\rm clad}$ from the lantern's geometric center, where $r_{\rm clad}$ is the radius of the lantern's cladding-jacket interface. However, by changing the placement of the outer ring (which can be achieved by etching down SMFs to different outer diameters, or inserting glass spacers into the fiber bundle before tapering), we can alter the 6-port lantern's response to modes such as defocus. Figure \ref{fig:osc}b plots defocus reconstruction range against lantern length for two kinds of non-selective 6-port lanterns: one where the outer cores are placed at a distance of $2/3 \, r_{\rm clad}$ from the center, and one where they are placed $1/2 \, r_{\rm clad}$ from the center. Changing the radial spacing of the lantern cores causes the oscillations in defocus reconstruction range to shift. Therefore, a 6-port lantern's linear range in the defocus mode can be tuned either by altering taper length or core spacing (though the former is likely easier to control). 
\\\\
Core arrangement also matters for hybrid lanterns. One trend we expect from radial symmetry arguments, and have confirmed through simulation, is that the selective core should not be placed in the lantern's center: doing so can remove sensitivity to circularly symmetric aberration modes. 

\subsection{Improved sensing with pre-lantern phase offsets}\label{ssec:masks}

\begin{figure}
    \centering
    \includegraphics[width=\textwidth]{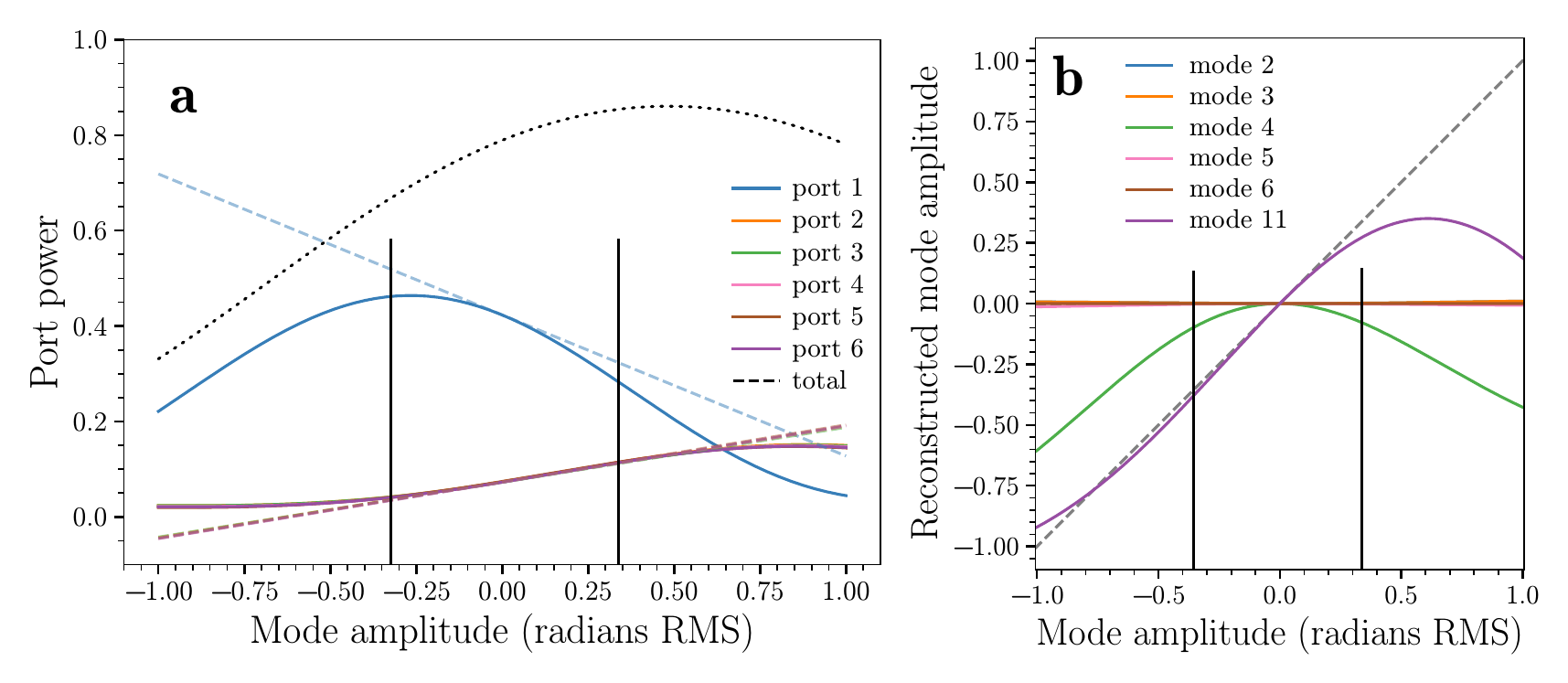}
    \caption{Panel a: intensity response of a standard 6-port lantern; a pupil-plane phase mask applies a static amount of spherical aberration (mode amplitude -0.5 radians) before injection into the PL. The dotted black curve tracks the total flux in all lantern outputs. Like in Figure \ref{fig:piaa}, the dashed, colored lines show the linearly-approximated WFS response, while vertical black bars denote linear range.  Panel b: linearly reconstructed spherical aberration mode amplitude as a function of true mode amplitude.}
    \label{fig:sphab}
\end{figure}

In this section, we consider how the manipulation of light before the lantern can change a lantern's sensing characteristics. One such method is to apply a phase offset (either using a dedicated mask in the pupil plane, or a DM). As a proof-of-concept, we show that phase offsets can increase the number of sensed modes. Our fiducial example is the standard 6-port lantern, which alone can sense the first 5 non-piston Zernike modes. We find that we can add a sixth sensed mode simply by adding a phase mask that resembles the desired mode; this static wavefront error biases the intensity response (e.g. Figure 4 from \cite{paper1}) of the PLWFS, changing the slopes and enabling the mode to be sensed in the linear regime. As a tradeoff, the increase in rank from the phase mask will come at the cost of throughput. We numerically verify this method by applying a spherical aberration phase offset (mode amplitude -0.5 radians) in the pupil plane of our 6-port PLWFS model; its intensity response in the presence of spherical aberration, as well as the linearly reconstructed mode amplitudes, are shown in Figure \ref{fig:sphab}. Comparing the throughput indicated by Figure \ref{fig:sphab}a ($\sim$0.8) to that of a standard 6-port lantern without a phase mask ($\sim$0.88, as per Table \ref{tab:standard}), we see that the cost of this added sensitivity to spherical aberration is a $\sim$10\% drop in throughput. We also note that the linear range of the sensor in this setup is limited by cross-talk between the defocus and spherical aberration modes. 

\subsection{Improved sensing with post-lantern beam-combination}\label{ssec:beamcombining}
\begin{figure}
    \centering
    \includegraphics[width=\textwidth]{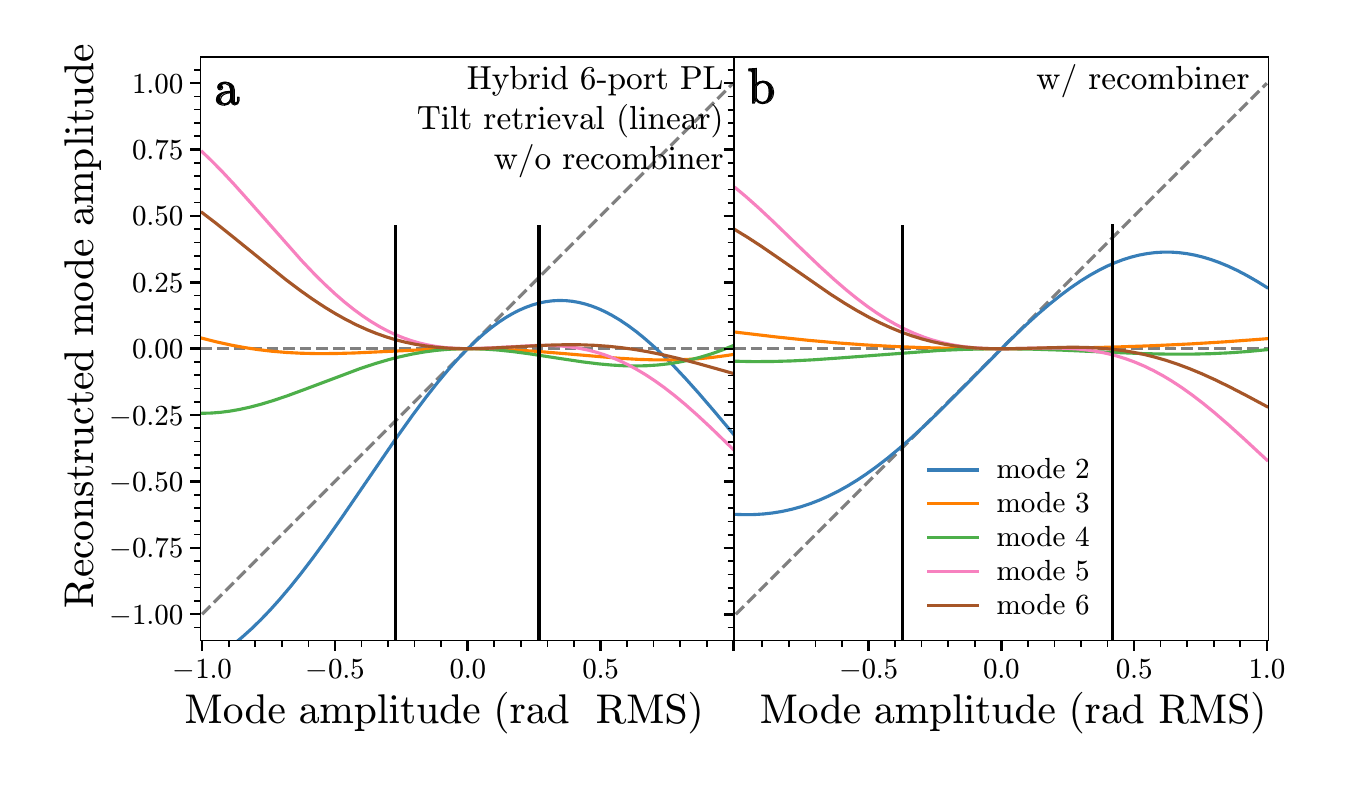}
    \caption{A beam recombiner, applied to the outputs of a 6-port hybrid PL, can increase the PL's linear reconstruction range. In this example, the simulated beam recombiner siphons 10\% of the light from the selective output and interferes it with each of the remaining ports, so that the 6 original outputs of the hybrid PL become 11 outputs.
    Panel a: reconstruction plot for tilt (Zernike mode 2), using a hybrid PL without the backend beam recombiner. Panel b: the same as a, but with the recombiner. The recombiner increases the reconstruction range for tilt by $\sim30\%-40\%$ }
    \label{fig:combiner1}
\end{figure}

Because PLs coherently couple light into single-mode cores, they naturally serve as a platform for the coherent manipulation of light through systems of beam splitters/combiners \cite{Diab:19}. These beam recombination systems have been produced within monolithic glass blocks using ultra-fast laser inscription (e.g. GLINT \cite{Martinod:21:GLINT}). Such devices can apply linear transformations to the complex-valued vector of PL outputs; as a result, compared to the phase mask approach of the previous section, we anticipate that beam recombiners will be able to more drastically change sensing characteristics, theoretically without loss of light. These transformations could be used to improve WFS capabilities, or to correct imperfections in lantern mode structure left by the manufacturing process. As a downside, this method requires more complicated, dedicated optics for complex beam-recombination. Such optics will likely be required for coherence-based exoplanet detection methods (Kim et al., submitted).
\\\\
As a proof-of-concept, we consider a simple beam recombination setup for PIAA-equipped hybrid lanterns, where the bulk of telescope light is coupled into a single ``selective'' output to be used for science. In this case, the PLWFS behaves more non-linearly, since all ports besides the science port now have near-zero intensity. To improve linearity, we can divert a small fraction $f$ of the light from the selective port, split it into $N-1$ equal parts, and pairwise-interfere each part with one of the $N-1$ non-selective ports, where $N$ is the number of PL outputs. For illustration purposes, we assume a unitary transfer matrix $A_{\rm 50:50}$ for each beam pair of the form
\begin{equation}\label{eq:split}
A_{\rm 50:50} = \dfrac{1}{\sqrt{2}}
\begin{pmatrix}
1 & 1 \\
-1 & 1
\end{pmatrix}.
\end{equation}
The above matrix corresponds to a ``50:50'' beam splitter; the overall beam recombination system turns $N$ PL outputs into $2N-1$ system outputs. Figure \ref{fig:combiner1} shows the effect of such a beam recombiner for a 6-port hybrid PL (the same as the one simulated in \S\ref{sec:results}, with beam-shaping optics), reconstructing tilt with the linear model. We find that this method trades throughput in the science port for linearity; the beam recombiner lowers flux in the selective port by 10\% but improves reconstruction range by $\sim30\%-40\%$. 
\\\\
In the future, we envision recombiner designs that improve sensing properties while preserving the number of PL outputs. Recombiners can also be constructed from more advanced building blocks such as tricouplers or photonic Mach-Zehnder interferometers; the latter architecture can enable the application of an arbitrary unitary transformation to the PL outputs \cite{reck}. These possibilities are further discussed in \S\ref{sec:disc}.

\section{Discussion}\label{sec:disc}
This work presents an initial exploration of lantern-based focal-plane wavefront sensors, with additional emphasis on contexts such as
high-resolution spectrometry and VFN where PLs already have 
strong arguments for application. In the process, we provide 
both an initial assessment of the PLWFS's capabilities, as well as preliminary investigations into optimization strategies. We discuss our 
results in each area, in turn. 

\subsection{PLWFS characterization}\label{ssec:PLWFSchar}
We observed several trends in the WFS metrics computed by our numerical model. First and foremost is that for both standard and hybrid lanterns, throughput and the number of sensed modes increase with number of ports --- expected, assuming that the number of output ports matches the number of accepted modes at the lantern entrance, as was done throughout this work. In such a scenario, an increase in the number of ports both increases the amount of light that can be captured, and decreases the amount of information lost in the process. 
We also found that hybrid lanterns exhibit a slightly higher degree of non-linearity over their standard counterparts, perhaps due to their more complicated mode structures. Linear reconstruction ranges of $	\lesssim 0.5$ radians RMS imply that PLWFSs will likely need to work in tandem with a first-stage pupil-plane WFS: these pupil-plane control schemes can remove larger WFEs (such as those from atmospheric turbulence) while leaving the correction of lower-amplitude residuals, produced by NCPAs and potentially even petaling modes, to the PLWFS. However, in the future these limitations may be lifted by exploiting spectral dispersion and polarization, employing more advanced non-linear reconstruction methods, and increasing mode counts.
\\\\
We also note that for all sensor configurations tested in \S\ref{ssec:allresults}, the number of sensed modes was strictly less than the number of outputs; we propose that this deficiency is due to a piston-like mode corresponding to the LP$_{01}$, backpropagated to the pupil plane, which takes up one degree of freedom but is not useful for wavefront sensing. This explanation is corroborated by \S\ref{ssec:masks}, where we showed that a 6-port PL could sense a full 6 Zernike modes though use of a pupil-plane phase mask, in essence replacing piston with a more useful mode. Beyond the aforementioned effect, we also expect to lose an additional degree of freedom in real systems because the sum of all sensor outputs can carry no WFS signature: there are many high-spatial-frequency aberration modes which will induce a drop in overall throughput. conversely, in simulations where we inject and retrieve exactly the same number of modes, and ignore confusion with higher-order modes, the sum of all ports can have WFS signature. Regardless of how degrees of freedom are lost, we may be able to gain them back by spectrally dispersing the PL outputs or exploiting polarization --- we leave this for future work. 
\\\\
We additionally test the impact beam-shaping optics and vortex masks have on the sensing performance of the PLWFS. Tests with the former are motivated by high-resolution fiber-fed spectrometry, where it is desirable to inject as much light as possible into the single diffraction-limited fiber that feeds the spectrometer, and then use the remaining light for wavefront sensing. We found that beam-shaping can route upwards of 80\% of total light into the selective core, at the cost of sensor linearity: the lack of light in the non-selective cores forces responses in those cores to behave more quadratically. When accounting for noise, the low fluxes in the non-selective ports will hinder wavefront sensing even more. We envision several ways to circumvent this issue, for instance by altering the PIAA lens design to leave some minimum amount of light in the non-selective ports, applying phase offsets with DMs as in \S\ref{ssec:masks}, or applying beam recombiners as in \S\ref{ssec:beamcombining}.
\\\\
Our tests with vortex masks are motivated by VFN. In this setup, we envision using a vortex mask and a PL to simultaneously perform VFN in one port and wavefront sensing/control with the other ports (for an alternative VFN setup using fully mode-selective lanterns, see \cite{Xin:22}). We find that the interaction between vortex masks and PLs is complex, alternately removing sensitivity to aberration modes, swapping sensitivity between modes, or adding sensitivity to a completely new mode. Interestingly, we also find that vortex masks can significantly increase the degenerate radius for standard PLs; heuristically, we believe this is due to the vortex mask introducing an asymmetry into the system. Further optimizations in this sort of setup will need to be made on a case-by-case basis, owing to the complex interaction between vortex masks and the PLWFS. In cases where the lantern modes have certain azimuthal symmetries (for instance, partially mode-selective lanterns), the impact of vortex masks may be more analytically tractable. 
\\\\
The next step in characterization is both an improvement in realism of these simulated lanterns (for instance, by modelling imperfections in the lantern manufacturing process, including non-circular claddings, or non-linear tapering profiles), and experimental verification, as well as an assessment of closed-loop performance and sensitivity in the presence of noise.

\subsection{Design optimization}\label{ssec:opt}
In this work, we considered how to improve the sensing performance of the PLWFS in two main areas: linear reconstruction range, and the number of sensed modes. One strategy to improve reconstruction range is to adjust lantern length: as in Figure \ref{fig:osc}, we found that reconstruction ranges for certain modes can oscillate from 0, marking complete insensitivity, to some maximal value as length is varied. Note that such a strategy is only possible if the manufacturing tolerances in PL length are small compared to the oscillation period ($\sim$ 300 $\upmu$m for our simulated 3-port and 6-port lanterns); alternatively, lantern length can be controlled to a lesser extent by etching back the lantern entrance post-fabrication, although care should be taken not to widen the lantern entrance so much that the shape or number of modes supported at the lantern entrance changes drastically.  The radial spacing between lantern cores also impacts wavefront sensing, as was shown in section \S\ref{ssec:coregeom}. In the future, we might envision even more exotic core arrangements with asymmetrically-placed cores. Because the parameter space of PL geometries is so large, the identification of ``good'' PL designs will likely require the use of waveguide mode-solvers working in tandem with numerical optimization techniques. 
\\\\
But improvement to PLWFS performance does not necessarily require alteration of the PL itself. Instead, we can manipulate light before and after the PL, through optics such as phase masks and beam recombiners. The first technique, considered in section \S\ref{ssec:masks}, involves applying a static phase offset in the pupil plane of the telescope-PLWFS system. In our example from \S\ref{ssec:masks}, we demonstrated that this technique could make a 6-port PL additionally sensitive to spherical aberration; however, this came at the cost of throughput, and the PLWFS's reconstruction range was ultimately limited by cross-talk between the spherical aberration and defocus modes. More refined phase masks, designed through global numerical optimization methods, may be able to mitigate these issues, and boost other properties like sensor linearity. Another natural continuation would be to consider masks (or dual-DM setups) that alter both amplitude and phase. We leave both ideas for future work. Phase offsets applied by DMs have the added benefit that they can be dynamically controlled, perhaps ramping down in amplitude as the overall wavefront error decreases, or switching on the fly to modify the set of sensed aberration mode as needed. 
\\\\
As shown in section \S\ref{ssec:beamcombining}, we can also improve PLWFS performance by using chains of beam splitters to repeatedly interfere the PL's outputs. Mathematically, this process applies a matrix transformation to the vector of complex SMF mode amplitudes at the lantern output, and can thereby alter the linearity of the system through the $Q$ metric (\S3.3 in \cite{paper1}). These transformations can additionally be designed to be norm-preserving so that no throughput losses will occur beyond those due to optical imperfections. As a proof-of-concept, we used a system of 50:50 beam splitters to improve the linear reconstruction range of a hybrid lantern; the next step will be to design more complex beam-combining systems. One promising architecture is the ``Mach-Zehnder'' mesh \cite{mzmesh} --- a photonic circuit composed of multiple stages of Mach-Zehnder interferometers which repeatedly interfere pairs of beams. Such a mesh, combined with piezoelectrically or thermally tunable phase delays \cite{Dong2022}, could enable active control of the PL outputs. Such a device could be operated to extend sensitivity to more aberration modes, increase the overall reconstruction range of the sensor, correct for PL manufacturing imperfections, and even function as a secondary stage of wavefront control.
\\\\
In all, there are four general classes of methods through which we can improve PLWFS performance: ``frontend'' methods like phase masks, which manipulate light before the PL; ``intrinsic'' methods which directly alter the lantern geometry; ``backend'' methods like beam-combining, which manipulate confined light after the PL; and finally, post-processing methods (e.g. the reconstruction model). All four are important in achieving greater control over the sensing properties of the PLWFS.

\section{Conclusion}
In this work, we used numerical models and the reconstruction methods from \cite{paper1} to benchmark the sensing performance of a variety of PLWFS configurations, covering both standard PLs and hybrid PLs, as well as the impact of beam-shaping optics in \S\ref{ssec:beamshaping} and vortex masks in \S\ref{ssec:vortex}. We find that, in conjunction with a hybrid lantern, beam-shaping can couple the bulk of incident light into the selective core, at the cost of a significant decrease of $\sim 40$\% in linear reconstruction range. The interaction between vortex masks and PLWFSs is less clear: alternatively adding to, changing, or subtracting from the set of sensed modes.
Results are tabulated in Tables \ref{tab:standard} and \ref{tab:hybrid}. In general, linear reconstruction ranges are found to be $\lesssim 0.5$ radians RMS, and degeneracies in the sensor response typically appear when WFE reaches 1-2 radians RMS. Our results do not mark a fundamental upper limit to the performance of the PLWFS: future pathways for improvement include spectral dispersion of lantern outputs, exploiting polarization, and switching from linear to more advanced non-linear and/or neural net reconstruction methods. We also consider how sensing performance could be improved by altering aspects of the PLWFS design, in \S\ref{sec:opt}.
\\\\
We find that improvements can be achieved either by optimizing the design of the lantern itself, or through the use of additional optics. We test two types of modifications to lantern design: adjustment of taper length and the radial spacing of lantern cores. We show that both parameters can strongly impact the reconstruction range for certain modes; in fact, we find that reconstruction range can oscillate with taper length and occasionally drop to 0, marking complete insensitivity to certain modes. Beyond modification of the lantern itself, we find that pupil-plane phase masks and backend beam recombiners can also increase sensing performance. In Section \S\ref{ssec:masks} we show that a standard 6-port lantern can additionally sense spherical aberration --- a mode that this lantern is insensitive to, without additional optics --- simply by applying a static spherical aberration WFE. Lastly, in \S\ref{ssec:beamcombining} we show that interferometric recombination of the output ports can improve sensor linearity.

%
\section*{Acknowledgements}

This material is based upon work supported by the National Science Foundation Graduate Research Fellowship Program under Grant No. DGE-2034835. Any opinions, findings, and conclusions or recommendations expressed in this material are those of the author(s) and do not necessarily reflect the views of the National
Science Foundation. This work was also supported by the National Science Foundation under Grant No. 2109232.

\section*{Disclosures}
The authors declare no conflicts of interest.
\appendix
\section{Types of photonic lanterns}\label{ap:a}
There are four types of PLs referenced throughout this work; to distinguish between them, we will consider the lantern's principal modes, the $N$ complex-valued distributions of light at the lantern entrance, each of which an $N$-port lantern couples into a single one of its outputs. Alternatively, the principal modes can be thought of as the front-end fields produced when reverse-injecting the PL through each of its SMF outputs. The prinicpal modes are occasionally referred to simply as ``lantern modes'', e.g. \cite{Leon-Saval:14,Lin:21}. 
\\\\
Differences can also be expressed through each lantern's complex-valued transfer matrix $U$, the unitary matrix that transforms the input vector expressing the electric field at the lantern entrance (in the basis of ``entrance modes'', the eigenmodes supported by the fiber-like lantern entrance) into a complex-valued vector of single-mode output amplitudes. In the limit where the single-mode cores taper to negligible size at the lantern entrance, and assuming circular symmetry, the entrance modes are the LP modes.

\paragraph{Non-selective or ``standard''}
A non-selective, non-mode-selective, or ``standard'' photonic lantern contains single-mode cores of uniform size and refractive index. As such, modes in any basis propagating through the lantern become degenerate in their propagation constants, leading to cross-coupling. The principal modes of a standard lantern are complex linear combinations of the entrance modes, and the transfer matrix $U$ has no clear properties other than being unitary.

\paragraph{Partially-selective or ``hybrid''}
In general, a ``hybrid'' or partially-selective PL ``selects'' a subset of the entrance modes, exclusively routing each to its own single-moded output; light in the remaining entrance modes is coupled non-selectively into the remaining outputs. The hybrid PL considered throughout this work and in \cite{hybrid} is specifically ``fundamental-mode-selective'', with only one core larger (or of higher index) than the rest. The output corresponding to this core exclusively couples all light from the fundamental entrance mode. The principal modes for the smaller ``non-selective'' cores are linear combinations of entrance modes, excluding the fundamental. The transfer matrix for an $N$-port hybrid PL that selects out only the fundamental mode has the block diagonal form of
\begin{equation}
    U = \begin{bmatrix}
        1 & 0 \\
        0 & U_{\rm N-1}
    \end{bmatrix}
\end{equation}
where $U_{\rm N-1}$ is an $N-1\times N-1$ unitary matrix. There can also be a phase variation along the diagonal, which we have omitted for clarity.
\paragraph{Mode-group-selective}
The entrance modes can be divided into groups sharing the same propagation constant. For instance, LP$_{01}$ is its own group, while the two LP$_{11}$ modes correspond to another. A mode-group-selective lantern uses differing core geometries to route each mode group to its own subset of single-mode outputs. The transfer matrix has the block diagonal form
\begin{equation}
    U = \begin{bmatrix}
        1 && 0\\
        & U_1\\
        0&& \ddots
    \end{bmatrix}
\end{equation}
where each $U_i$ is an $N_i\times N_i$ unitary matrix corresponding to mode group $i$ with $N_i$ modes. 
\paragraph{Mode-selective}
A mode-selective or ``fully-mode-selective'' lantern uses differing core geometries {\it and} asymmetries in the waveguide structure to break all degeneracies between entrance modes, and then route each entrance mode to a separate output. The principal modes of a mode-selective lantern are the entrance modes. The transfer matrix of an ideal lantern of this type is identity:
\begin{equation}
    U = \begin{bmatrix}
        1 & 0 \\
        0 & \ddots 
    \end{bmatrix}
\end{equation}
The 3-port lantern from \cite{Leon-Saval:14}, which has one large core and two small cores, and an asymmetric cladding shape, is fully mode-selective. The same lantern with a circular cladding, like the one simulated in this work, is mode-group-selective, and also hybrid.

\section{Clarification: solving for degenerate radius} \label{ap:deg}
As defined in \cite{paper1}, the degenerate radius is the minimum amount of RMS WFE required to zero a column of the Jacobian: the matrix of partial derivatives of port flux with respect to aberration amplitude. This particular criterion was chosen as a proxy for degeneracy in order to enable the use of numerical root-finding techniques. We note a minor correction to the formula provided in \cite{paper1}. Assume a modal basis for phase aberrations with a change-of-basis matrix $R$, such that a phase map $\bm{\phi}$ can be expressed as a mode amplitude vector $\bm{a}$, with $\bm{\phi} \equiv R \bm{a}$. In this basis, the Jacobian evaluated about $\bm{\phi}$ is
\begin{equation}
    \dfrac{\partial (p_{{\rm out},i}/p_{\rm in})}{\partial a_k} = - 2 \, {\rm{ Im}} \left[ \sum_{j} A_{ij}e^{i \phi_j} R_{jk}  \times \sum_{j'} A^*_{ij'}e^{-i\phi_{j'}}   \right].
\end{equation}
where $p_{{\rm out},i}/p_{\rm in}$ is the flux in the $i$th lantern output normalized by total incident flux, $A_{ij}$ is the complex-valued transfer matrix of the lantern in pixel basis, and {\rm Im} denotes the imaginary part. Note that both the new formula and the old, which had an extra column-wise multiplication, will give the same value for degenerate radius.

\bibliography{refs}

\end{document}